\newif\ifOneCol

\OneColtrue
\OneColfalse

\ifOneCol
\documentclass[12pt, onecolumn, draftclsnofoot, lettersize, journal]{IEEEtran}
\else
\documentclass[lettersize, journal]{IEEEtran}
\fi
\IEEEoverridecommandlockouts
\usepackage[hidelinks]{hyperref}
\usepackage{orcidlink}
\usepackage{cite}
\usepackage{amsmath,amssymb,amsfonts}
\usepackage{algorithmic}
\usepackage{graphicx}
\usepackage{textcomp}
\usepackage{xcolor}
\def\BibTeX{{\rm B\kern-.05em{\sc i\kern-.025em b}\kern-.08em
    T\kern-.1667em\lower.7ex\hbox{E}\kern-.125emX}}
\usepackage[nolist,nohyperlinks]{acronym}
\usepackage{booktabs}
\usepackage{siunitx}
\usepackage{tikz}
\usepackage{etoolbox} % for \ifnumcomp
\usepackage{listofitems} % for \readlist to create arrays
\usepackage{amsmath}
\usepackage[utf8]{inputenc}
\usepackage{pgfplots} 
\usepackage{pgfgantt}
\usepackage{pdflscape}
\pgfplotsset{compat=newest} 
\pgfplotsset{plot coordinates/math parser=false}

\tikzset{>=latex} % for LaTeX arrow head
\colorlet{myred}{red!80!black}
\colorlet{myblue}{blue!80!black}
\colorlet{mygreen}{green!60!black}
\colorlet{mydarkred}{myred!40!black}
\colorlet{mydarkblue}{myblue!40!black}
\colorlet{mydarkgreen}{mygreen!40!black}
\tikzstyle{node}=[very thick,circle,draw=myblue,minimum size=12,inner sep=0.5,outer sep=0.6]
\tikzstyle{connect}=[->,thick,black,shorten >=1]
\tikzset{ % node styles, numbered for easy mapping with \nstyle
  node 1/.style={node,mydarkgreen,draw=mygreen,fill=mygreen!25},
  node 2/.style={node,mydarkblue,draw=myblue,fill=myblue!20},
  node 3/.style={node,mydarkred,draw=myred,fill=myred!20},
}
\def\nstyle{int(\lay<\Nnodlen?min(2,\lay):3)} % map layer number onto 1, 2, or 3

\usepackage{comment}
\DeclareMathAlphabet\mathbfcal{OMS}{cmsy}{b}{n}
\newcolumntype{C}[1]{>{\centering\let\newline\\\arraybackslash\hspace{0pt}}m{#1}}

\makeatletter
\def\ps@headings{%
\def\@oddhead{\parbox[t][\height][t]{\textwidth}{\centering \scriptsize
This work has been submitted to the IEEE for possible publication. \\Copyright may be transferred without notice, after which this version may no longer be accessible.
}\hfil\hbox{}}%

\def\@evenhead{\parbox[t][\height][t]{\textwidth}{\centering \scriptsize
This work has been submitted to the IEEE for possible publication. \\Copyright may be transferred without notice, after which this version may no longer be accessible.
}\hfil\hbox{}}%
}

\def\ps@IEEEtitlepagestyle{%
\def\@oddhead{\parbox[t][\height][t]{\textwidth}{\centering \scriptsize
This work has been submitted to the IEEE for possible publication. \\Copyright may be transferred without notice, after which this version may no longer be accessible.
}\hfil\hbox{}}%
\def\@evenhead{ \hfil \leftmark\mbox{}}%
}

\makeatother
\begin{document}
\bstctlcite{IEEEexample:BSTcontrol}

\title{A Low-Complexity Machine Learning Design for mmWave Beam Prediction\\
	}
\author{{Muhammad Qurratulain Khan\,\orcidlink{0000-0001-5779-0101}},
{Abdo Gaber\,\orcidlink{0000-0001-5779-0101}},
{Mohammad Parvini\,\orcidlink{0000-0002-1315-7635}}, \IEEEmembership{Member,~IEEE},
{Philipp Schulz\,\orcidlink{0000-0002-0738-556X}},
{and Gerhard Fettweis\,\orcidlink{0000-0003-4622-1311}}, \IEEEmembership{Fellow, IEEE}
\thanks{This work was funded by the European Union's SEMANTIC ITN project under the Marie Skodowska-Curie grant agreement No. 861165.}
\thanks{Muhammad Qurratulain Khan and Abdo Gaber are with the National Instruments Corporation, Dresden, Germany. (email: \{muhammad.qurratulain.khan, abdo.nasser.ali.gaber\}@ni.com).}
\thanks{Mohammad Parvini, Philipp Schulz, and Gerhard Fettweis are with the Vodafone Chair for Mobile Communications Systems, Technische Universität Dresden, Germany. (email: \{mohammad.parvini, philipp.schulz2, gerhard.fettweis\}@tu-dresden.de).}
}
	\maketitle
\begin{acronym}
    \acro{mmWave}{millimeter-wave}
    \acro{5G}{fifth generation}
    \acro{THz}{terahertz}
    \acro{BS}{base station}
    \acro{UE}{user equipment}
    \acro{3GPP}{3rd Generation Partnership Project} 
    \acro{SSB}{synchronization sequence block}
    \acro{RSRP}{reference signal received power}
    \acro{EBS}{exhaustive beam scan}
    \acro{HBS}{hierarchical beam scan}
    \acro{AI}{artificial intelligence}
    \acro{ML}{machine learning}
    \acro{FC}{fully-connected}
    \acro{NN}{neural network}
    \acro{DNN}{deep neural network}
    \acro{CNN}{convolutional neural network}
    \acro{NR}{New Radio}
    \acro{MIMO}{multi-in multi-out}
    \acro{LOS}{line-of-sight}
    \acro{NLOS}{non-line-of-sight}
    \acro{AoA}{angle-of-arrival}
    \acro{AoD}{angle-of-departure}
    \acro{UPA}{uniform planar array}
    \acro{RF}{radio frequency}
    \acro{AWGN}{additive white Gaussian noise}
    \acro{KPI}{key performance indicator}
    \acro{CDL}{clustered delay line}
    \acro{FLOP}{floating-point operation}
\end{acronym}
%TC:endignore
\begin{abstract}
The \ac{3GPP} is currently studying \ac{ML} for the \ac{5G}-Advanced \ac{NR} air interface, where spatial and temporal-domain beam prediction are important use cases. With this background, this letter presents a low-complexity \ac{ML} design that expedites the spatial-domain beam prediction to reduce the power consumption and the reference signaling overhead, which are currently imperative for frequent beam measurements. Complexity analysis and evaluation results showcase that the proposed model achieves state-of-the-art accuracy with lower computational complexity, resulting in reduced power consumption and faster beam prediction. Furthermore, important observations on the generalization of the proposed model are presented~in~this~letter.
\end{abstract}
%TC:ignore
\acresetall
\section{Introduction}
The availability of abundant bandwidth at \ac{mmWave} bands makes it a requisite for higher throughput. However, to achieve an adequate link margin, beamforming via large antenna arrays is essential \cite{7109864}. Consequently, the evaluation of beam qualities through frequent beam measurements and beam qualities reporting is imperative to help the \ac{BS} and the \ac{UE} decide the optimal beam pair for link establishment. Within the \ac{3GPP} this is referred to as beam management procedure.

In order to enable the \ac{UE} to measure the beam qualities, beamformed reference signals (\acp{SSB}) are sequentially transmitted from the \ac{BS} in the form of an SSB burst. This allows the UE to measure the qualities of all the \ac{BS} transmit beams in terms of their \acp{RSRP} through one of its receive beams. Further, to measure the qualities of all possible transmit-receive beam pairs, several SSB bursts are transmitted. This procedure of beam qualities measurement is known as \ac{EBS}, which suffers from large beam measurement overhead, increased latency, and higher power consumption \cite{10036372, 10123939}. To overcome this, a two-level \ac{HBS} consisting of parent (wide) and child (narrow) beams is employed \cite{7390101}. Nevertheless, it suffers from increased latency and inaccuracy of beam~selection.

Recently, \ac{ML} methods have been extensively applied to wireless communications to solve the non-linear problems that were burdensome to be resolved by conventional signal processing techniques. Consequently, several studies propose the use of \ac{ML} for beam prediction and selection \cite{10036372}. A straightforward approach to reduce the beam measurement overhead is to utilize the \ac{UE} location information \cite{9013296} to train an \ac{ML} model for beam prediction. However, transmission of \ac{UE} location information, which may not necessarily be available always to the \ac{BS}, poses an additional feedback overhead. To avoid this issue, the study in \cite{9314253} fuses the concept of \ac{HBS} with a supervised \ac{ML} model and exploits the spatial correlation among the parent and the child beam qualities to predict the optimal child beam. A similar approach in \cite{9492142} utilizes the received signal vector of parent beams as an input to a \ac{CNN}. Another approach in \cite{10001224} proposes to reduce the beam measurement overhead by transmitting a subset of child beams and then utilizes a \ac{CNN} that predicts the optimal beam by learning the spatial correlation among child beams.

Starting from 2022, the study of \ac{ML} for the \ac{5G}-Advanced \ac{NR} air interface is an important project at \ac{3GPP}. Here, the focus is to explore the benefits of augmenting the \ac{NR} air interface with \ac{ML} models for enhanced performance and/or reduced overhead and complexity \cite{3GPP1}. An important study item in this project is the evaluation of \ac{ML} for beam management, where spatial and temporal-domain beam prediction are the sub use cases \cite{3GPP2}. Following \ac{3GPP} guidelines, companies report their proposed evaluation methodology and results on \ac{ML}-based beam prediction \cite{3GPP3}. A recent proposal for spatial-domain beam prediction is presented in \cite{3GPP4}, where based on the received power of a subset of the transmit beams, a \ac{CNN} is trained to predict the \acp{RSRP} of the non-transmitted beams resulting in reduced overhead.

Though most of the discussed \ac{ML} solutions reduce the beam measurement overhead while achieving a performance closer to \ac{EBS}, no significant attention has been paid to the model computational complexity, model training time and its generalization capabilities. To bridge this research gap, this letter presents a low-complexity \ac{ML} beam prediction approach that achieves the performance closer to the optimal \ac{EBS} but with lower computational complexity as compared to other ML approaches, resulting in faster beam prediction. Additionally, to investigate the generalization capabilities of our model, we evaluate its performance over \ac{3GPP} specified scenarios.
\section{System Model}
This section details channel and beam steering models, followed by an overview of the beam management procedure.
\subsection{Channel Model}
We consider a downlink \ac{mmWave} \ac{MIMO} communication system, where the \ac{BS} and the \ac{UE} are equipped with $N_{\mathrm{T}}$ and $N_{\mathrm{R}}$ antenna elements, respectively. Using the clustered channel model, the channel is assumed to be the sum of the \ac{LOS} path and $C$ \ac{NLOS} clusters with $L$ paths per cluster. The channel matrix $\textbf{H} \in \mathbb{C}^{N_{\mathrm{R}} \times N_{\mathrm{T}}}$ can then be written as \cite{3GPP5}
\ifOneCol
\begin{multline}\label{channelModel}
\textbf{H} = \underbrace {\sqrt{\frac{K \Lambda}{K+1}}  \alpha_{\text{LOS}}
\textbf{a}_{\mathrm{R}}(\phi^{\mathrm{R}}_{\text{LOS}},\theta^{\mathrm{R}}_{\text{LOS}}) \textbf{a}_{\mathrm{T}}^{H}(\phi^{\mathrm{T}}_{\text{LOS}},\theta^{\mathrm{T}}_{\text{LOS}})}_{\textbf{H}_{\text{LOS}}}
\\
+ 
\underbrace {\sqrt{\frac{\Lambda}{L(K+1)}}
\sum_{c=1}^{C} \sum_{l=1}^{L} \alpha_{c,l}
\textbf{a}_{\mathrm{R}}(\phi^{\mathrm{R}}_{c,l},\theta^{\mathrm{R}}_{c,l}) \textbf{a}_{\mathrm{T}}^{{H}}(\phi^{\mathrm{T}}_{c,l},\theta^{\mathrm{T}}_{c,l})}_{\textbf{H}_{\text{NLOS}}}.
\end{multline}
% One Column Version
%
\else
\begin{multline}\label{channelModel}
\textbf{H} = \underbrace {\sqrt{\frac{K \Lambda}{K+1}}  \alpha_{\text{LOS}}
\textbf{a}_{\mathrm{R}}(\phi^{\mathrm{R}}_{\text{LOS}},\theta^{\mathrm{R}}_{\text{LOS}}) \textbf{a}_{\mathrm{T}}^{H}(\phi^{\mathrm{T}}_{\text{LOS}},\theta^{\mathrm{T}}_{\text{LOS}})}_{\textbf{H}_{\text{LOS}}}
\\
+ 
\underbrace {\sqrt{\frac{\Lambda}{L(K+1)}}
\sum_{c=1}^{C} \sum_{l=1}^{L} \alpha_{c,l}
\textbf{a}_{\mathrm{R}}(\phi^{\mathrm{R}}_{c,l},\theta^{\mathrm{R}}_{c,l}) \textbf{a}_{\mathrm{T}}^{{H}}(\phi^{\mathrm{T}}_{c,l},\theta^{\mathrm{T}}_{c,l})}_{\textbf{H}_{\text{NLOS}}}.
\end{multline}
% two column version
\fi

Here, the $l$-th path of the $c$-th cluster has azimuth (elevation) \ac{AoA} $\phi^{\mathrm{R}}_{c,l} (\theta^{\mathrm{R}}_{c,l})$ and azimuth (elevation) \ac{AoD} $\phi^{\mathrm{T}}_{c,l} (\theta^{\mathrm{T}}_{c,l})$, while $\alpha_{c,l}$ is the complex path gain. The same variables are analogously defined for the \ac{LOS} path and are indicated by the \ac{LOS} index. Furthermore, $\textbf{a}_{\mathrm{R}}(\cdot) \in \mathbb{C}^{N_{\mathrm{R}} \times 1}$ and $\textbf{a}_{\mathrm{T}}(\cdot)  \in \mathbb{C}^{N_{\mathrm{T}} \times 1}$ denote the \ac{UE} and the \ac{BS} array response, respectively, $(\cdot)^{{H}}$ denotes conjugate transpose, $K$ is the Ricean factor, and $\Lambda$ indicates the pathloss.

We assume a \ac{UPA} in the $\mathrm{y}$-$\mathrm{z}$ plane at the \ac{BS} and the \ac{UE} with $N_{\mathrm{y}}$ and $N_{\mathrm{z}}$ antenna elements ($N_{\mathrm{y}}N_{\mathrm{z}} = N$) on $\mathrm{y}$ and $\mathrm{z}$ axis, respectively. Here, for ease of notation we drop the subscript for the \ac{BS} and \ac{UE}. The array response vector for the \ac{UPA} can then be written as
\ifOneCol
\begin{equation}\label{arrayResponse}
    \textbf{a}(\phi,\theta) = \frac{1}{\sqrt{N}}[1,  \cdots,  e^{j \frac{2 \pi}{\lambda} d (y'\sin(\phi)\sin(\theta)+z'\cos(\theta))}, \cdots, 
    e^{j \frac{2 \pi}{\lambda} d ((N_{\mathrm{y}}-1)\sin(\phi)\sin(\theta)+(N_{\mathrm{z}}-1)\cos(\theta))}  ]^{{T}},
\end{equation}
\else
\begin{multline}\label{arrayResponse}
    \textbf{a}(\phi,\theta) = \frac{1}{\sqrt{N}}[1,  \cdots,  e^{j \frac{2 \pi}{\lambda} d (y'\sin(\phi)\sin(\theta)+z'\cos(\theta))}, \cdots, 
    \\
    e^{j \frac{2 \pi}{\lambda} d ((N_{\mathrm{y}}-1)\sin(\phi)\sin(\theta)+(N_{\mathrm{z}}-1)\cos(\theta))}  ]^{{T}},
\end{multline}
\fi
where $y' \in \{0, 1, \cdots, N_{\mathrm{y}}-1 \}$, $z' \in \{0, 1, \cdots, N_{\mathrm{z}}-1 \}$, while $ \lambda $ and $d=\frac{\lambda}{2}$ indicate the wavelength and antenna element spacing, respectively.
\subsection{Beam Steering Model}
We consider phase shifter based analog beamforming with one \ac{RF} chain. At the \ac{BS} the transmit signal is beamformed by a beamforming vector $\textbf{f} = [{f}_{1}, {f}_{2}, \cdots ,{f}_{N_{\mathrm{T}}}]^{{T}}$$\in\mathbb{C}^{N_{\mathrm{T}} \times 1}$ and at the \ac{UE} the received signals are combined with a receive combining vector $\textbf{w} = [{w}_{1}, {w}_{2}, \cdots ,{w}_{N_{\mathrm{R}}}]^{{T}}$$\in \mathbb{C}^{N_{\mathrm{R}} \times 1}$. Here, $f_{i}$ and $w_{j}$ denote the complex weight on the $i$-th transmit and $j$-th receive antenna element, respectively. The transmit and receive beams are selected from the predefined codebooks $\mathcal{F}$ and $\mathcal{W}$, consisting of $F$ and $W$ candidate beams, respectively. The codebooks are designed on the following beam steering scheme.
\begin{align}\label{codebookDesign}
    \textbf{f} \in \mathcal{F} &=   \{ \textbf{a}_{\mathrm{T}}(\Bar{\phi}^{\mathrm{T}}_{1},\Bar{\theta}^{\mathrm{T}}_{1}), \textbf{a}_{\mathrm{T}}(\Bar{\phi}^{\mathrm{T}}_{2},\Bar{\theta}^{\mathrm{T}}_{2}), \cdots ,
    \textbf{a}_{\mathrm{T}}(\Bar{\phi}^{\mathrm{T}}_{F},\Bar{\theta}^{\mathrm{T}}_{F})  \}
    \\
    \textbf{w} \in \mathcal{W} &=   \{ \textbf{a}_{\mathrm{R}}(\Bar{\phi}^{\mathrm{R}}_{1},\Bar{\theta}^{\mathrm{R}}_{1}), \textbf{a}_{\mathrm{R}}(\Bar{\phi}^{\mathrm{R}}_{2},\Bar{\theta}^{\mathrm{R}}_{2}), \cdots ,
    \textbf{a}_{\mathrm{R}}(\Bar{\phi}^{\mathrm{R}}_{W},\Bar{\theta}^{\mathrm{R}}_{W})  \}
\end{align}
Here, $\Bar{\phi}_{m}^{\mathrm{T}}(\Bar{\theta}_{m}^{\mathrm{T}})$ for the $m$-th transmitting beam $\textbf{f}_{m}$, $m \in \{1, 2, \cdots,  F\}$ and $\Bar{\phi}_{n}^{\mathrm{R}}(\Bar{\theta}_{n}^{\mathrm{R}})$ for the $n$-the receiving beam $\textbf{w}_{n}$, $n \in \{1, 2, \cdots,  W\}$ are the quantized azimuth (elevation) \ac{AoD} and \ac{AoA}, respectively.
Given the channel matrix $\textbf{H}$, the transmit signal $x$, the $m$-th transmitting beam $\textbf{f}_{m}$ and the $n$-th receiving beam $\textbf{w}_{n}$, the received signal $y_{m,n}$ is
\begin{equation}\label{receivedSignal}
    y_{m,n} = \sqrt{P}\textbf{w}_{n}^{H}\textbf{H}\textbf{f}_{m}x + \textbf{w}_{n}^{H}\boldsymbol{\eta},
\end{equation}
where ${P}$ is the transmit power and $\boldsymbol{\eta} \in \mathbb{C} ^ {{N}_{\mathrm{R}} \times 1}$ is the \ac{AWGN}.
\subsection{Beam Management in 5G NR}
The \ac{3GPP} beam management procedure is based on the \ac{EBS} and aims to find the optimal beam pair $\{\textbf{f}_{{m}^{*}}, \textbf{w}_{{n}^{*}} \}$ that maximizes the \ac{RSRP} given as: RSRP$_{m,n} = |y_{m,n}|^{2}$. The optimization problem can be formulated as 
\begin{equation}\label{EBS}
    \{m^{*}, n^{*} \} = \underset{{\substack{m\in \{1,2, \cdots, F \},   \\n\in \{1,2, \cdots, W \}}}}
    {\mathrm{argmax}}\, \text{RSRP}_{m,n}.
\end{equation}
\ac{EBS} solves this optimization problem by exhaustively searching over all possible beamforming and combining vectors leading to an excessively huge beam training overhead of $F \cdot W$ beam measurements.

To reduce this beam measurement overhead, \ac{HBS} utilizes a multi-resolution codebook and the problem of beam selection is divided into two levels. The first-level search identifies the best parent beam by solving
\begin{equation}\label{HBS-1}
    \{m^{*}_{\mathrm{p}}, n^{*}_{\mathrm{p}} \} = \underset{{\substack{m_{\mathrm{p}}\in \{1,2, \cdots, F_{\mathrm{p}} \},   \\n_{\mathrm{p}}\in \{1,2, \cdots, W_{\mathrm{p}} \}}}}
    {\mathrm{argmax}}\, \text{RSRP}^{\mathrm{p}}_{m_{\mathrm{p}},n_{\mathrm{p}}}.
\end{equation}
Here, $F_{\mathrm{p}} = \frac{F}{s_{\mathrm{T}}}$ and $W_{\mathrm{p}} = \frac{W}{s_{\mathrm{R}}}$ indicate the number of parent beams at the \ac{BS} and \ac{UE}, respectively. Further, $s_{\text{T}}$ and $s_{\text{R}}$ defines the number of child beams within each parent beam at the \ac{BS} and \ac{UE}, respectively. 
After identifying the best parent beam pair, the second-level search confirms the optimal child beam pair within the range of the selected parent beam~pair~\eqref{HBS-1},~by
\begin{equation}\label{HBS-2}
    \{m^{*}, n^{*} \} = \underset{{\substack{m\in \{(m^{*}_{\mathrm{p}}-1)s_{\mathrm{T}}+1, \cdots, m^{*}_{\mathrm{p}}s_{\mathrm{T}} \},   \\n\in \{(n^{*}_{\mathrm{p}}-1)s_{\mathrm{R}}+1, \cdots, n^{*}_{\mathrm{p}}s_{\mathrm{R}} \}}}}
    {\mathrm{argmax}}\, \text{RSRP}^{\mathrm{c}}_{m,n}.
\end{equation}
Notably, the first and the second-level search requires $F_{\mathrm{p}} \cdot W_{\mathrm{p}}$ and $s_{\mathrm{R}} \cdot s_{\mathrm{T}}$ beam measurements, respectively, resulting in reduced beam measurement overhead. However, the multi-level search incurs increased latency.
\section{Low-Complexity \acl{ML} Design for \ac{mmWave} Beam Prediction}
In this section, we leverage the angular domain spatial correlation to propose a low-complexity beam prediction model for fast beam training. Motivated by the fact that very large antenna arrays can only be employed at the \ac{BS} due to size constraints, in the following sections, we limit our discussion to the identification of the optimal transmit beam, i.e., the assumption of the knowledge of the optimal receive beam.
\ifOneCol
\begin{figure}[t]
	\centering
	\includegraphics{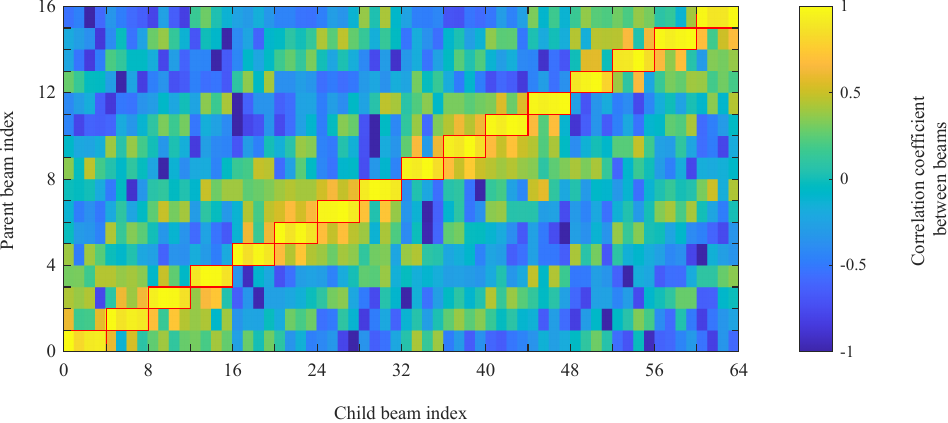}
	\caption{Spatial correlation among \acp{RSRP} of parent and child beams.}%, where each parent beam contains 4 child beams.}
	\label{fig:correlation}
\end{figure}
\fi
\subsection{Algorithm Framework}
Motivated by the two-level beam search, we propose to cover the whole angular region with the first-level parent beams. By doing so, we observe that there exists a strong angular spatial correlation among parent and child beams in a certain environment. As an example, Fig. \ref{fig:correlation} shows the angular spatial correlation between the \acp{RSRP} of the parent and the child beams, where each parent beam contains four child beams. Here, it can be observed that the parent beam has a stronger correlation with a limited number of child beams. Consequently, we assume that the $\text{RSRP}^{\mathrm{c}}$ of the child beams is a function $f_{1}(\cdot)$ of the parental \ac{RSRP} values, i.e.,
\begin{equation}\label{rsrpParentChild}
    \text{RSRP}^{\mathrm{c}} = f_{1}(\text{RSRP}^{\mathrm{p}}).
\end{equation}
In particular, we aim on probing the parent beams and obtaining their corresponding \acp{RSRP} from the received signal vector $\textbf{y}^{\mathrm{p}} = [y_{1}^{\mathrm{p}}, y_{2}^{\mathrm{p}}, \cdots, y_{F_{\mathrm{p}}}^{\mathrm{p}}]^{{T}}$ and by intelligently merging these parent \acp{RSRP} with the strong correlation among parent and child beams, we can predict the optimal child beam index $m^{*}$. Due to the discrete number of candidate beams, the beam prediction problem can be formulated as multiclass-classification problem and can be written as
\begin{equation}\label{rsrpOptIndex}
    m^{*} = f_{2}(\text{RSRP}^{\mathrm{p}}), \quad m^{*} \in \{1,2,\cdots,F\}
\end{equation}
where $f_{2}(\cdot)$ is the function that learns the correlation between parent and child \acp{RSRP} for optimal beam index prediction. Further, due to the highly non-linear relationship between \acp{RSRP} and channel directivity, the prediction is difficult to be estimated by conventional signal processing methods. With this background, we propose a low-complexity \ac{ML} design for beam prediction in the following section.
\ifOneCol
\else
\begin{figure}[t]
	\centering
	\includegraphics{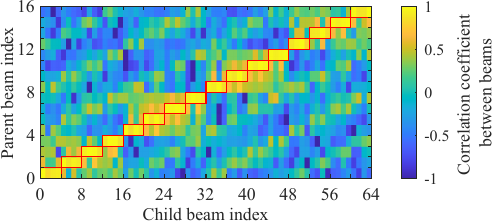}
	\caption{Spatial correlation among \acp{RSRP} of parent and child beams.}%, where each parent beam contains 4 child beams.}
	\label{fig:correlation}
\end{figure}
\fi
\ifOneCol
\begin{figure}[t]
	\centering
% NEURAL NETWORK
\begin{tikzpicture}[x=2cm,y=0.81cm]
  \readlist\Nnod{3,4} % array of number of nodes per layer
  \readlist\Nstr{F_{\mathrm{p}},m,F} % array of string number of nodes per layer
  \readlist\Cstr{\text{RSRP}^{\mathrm{p}},h^{(\prev)},\hat{\mathcal{P}}} % array of coefficient symbol per layer
  \def\yshift{0.5} % shift last node for dots
  % LOOP over LAYERS
  \foreachitem \N \in \Nnod{
    \def\lay{\Ncnt} % alias of index of current layer
    \pgfmathsetmacro\prev{int(\Ncnt-1)} % number of previous layer
    \foreach \i [evaluate={\c=int(\i==\N); \y=\N/2-\i-\c*\yshift;
                 \x=\lay; \n=\nstyle;
                 \index=(\i<\N?int(\i):"\Nstr[\n]");}] in {1,...,\N}{ % loop over nodes
      % NODES
      \node[node \n] (N\lay-\i) at (\x,\y) {};
      % CONNECTIONS
      \ifnumcomp{\lay}{>}{1}{ % connect to previous layer
        \foreach \j in {1,...,\Nnod[\prev]}{ % loop over nodes in previous layer
          \draw[white,line width=1,shorten >=1] (N\prev-\j) -- (N\lay-\i);
          \draw[connect] (N\prev-\j) -- (N\lay-\i);
        }
        \ifnum \lay=\Nnodlen
          \draw[connect] (N\lay-\i) --++ (0.5,0); % arrows in; % arrows out
          \draw[connect] (N\lay-\i) --++ (1.5,0) node[right, above , text = black] {$\strut\Cstr[\n]_{\index}$}; % arrows in; % arrows out
        \fi
      }{
        \draw[connect] (0.2,\y) -- (N\lay-\i) node[above, midway, text = black] {$\strut\Cstr[\n]_{\index}$}; % arrows in
      }
      
    }
    \path (N\lay-\N) --++ (0,1+\yshift) node[midway,scale=1] {$\vdots$}; % dots
  }
  % LABELS
 \node[rectangle,draw = myblue,text = black,fill = myblue!20,minimum width = 22, minimum height = 90] (r) at (2.81,-0.75) {Softmax};
\end{tikzpicture}
\caption{ Proposed low-complexity \ac{ML} design for beam prediction.}
\label{fig:NNModel}
\end{figure}
\fi
\subsection{Model Design}
In this section, we introduce our \ac{ML} model and its corresponding inputs and outputs as shown in Fig. \ref{fig:NNModel}. 
\subsubsection{Input Layer}
Based on our previous discussions, the \ac{RSRP}$^{\mathrm{p}}$ of the parent beams obtained via the first level of traditional \ac{HBS} is provided as an input to the model. This indicates that the input layer consists of $F_{\mathrm{p}}$ nodes. As an example, considering $F = 64$ beams and selecting $s_{\mathrm{T}} = 4$ results in $F_{\mathrm{p}}=16$ parent beams which means that a beam measurement overhead reduction of $1-\frac{16}{64} = 75\%$ is achieved as compared to the \ac{EBS}.
\subsubsection{Output Layer}
For the prediction of the optimal child beam from all the candidate child beams, a \ac{FC} layer, consisting of $F$ nodes is introduced, which learns the spatial correlation between \ac{RSRP}$^{\mathrm{p}}$ and \ac{RSRP}$^{\mathrm{c}}$ and transforms it to the candidate child beams. Finally, a non-linear softmax activation layer is introduced that returns the probabilities of all the child beams. The output of the proposed low-complexity \ac{NN} can be written as
\begin{equation}\label{softmax}
    \hat{\mathbfcal{P}} = \text{softmax}(\textbf{A}^{{T}}\text{\ac{RSRP}}^{\mathrm{p}}+\textbf{b}).
\end{equation}
Here, $\hat{\mathbfcal{P}} \in \mathbb{C} ^ {F \times 1}$ is the predicted output probability vector of all the child beams, while $\textbf{A} \in \mathbb{C} ^ {F_{\mathrm{p}} \times F}$ and $\textbf{b}\in \mathbb{C} ^ {F \times 1}$ are the weights and the biases, respectively. Finally, the child beam with maximum probability $\hat{\mathcal{P}}_{m}$ is selected, i.e., 
\begin{equation}\label{childBeam Selection}
    \hat{m}^{*}  = \underset{{\substack{m\in \{1,2, \cdots, F \}}}}
    {\mathrm{argmax}}\, \hat{\mathcal{P}}_{m}.
\end{equation}
\ifOneCol
\else
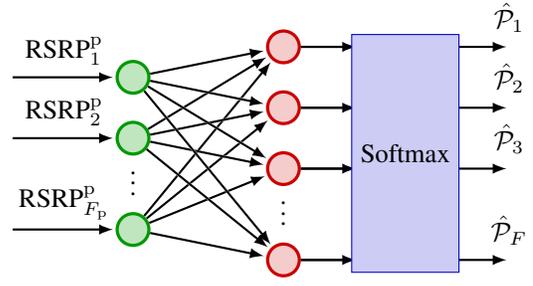
\begin{figure}[t]
	\centering
% NEURAL NETWORK
\begin{tikzpicture}[x=2cm,y=0.81cm]
  \readlist\Nnod{3,4} % array of number of nodes per layer
  \readlist\Nstr{F_{\mathrm{p}},m,F} % array of string number of nodes per layer
  \readlist\Cstr{\text{RSRP}^{\mathrm{p}},h^{(\prev)},\hat{\mathcal{P}}} % array of coefficient symbol per layer
  \def\yshift{0.5} % shift last node for dots
  % LOOP over LAYERS
  \foreachitem \N \in \Nnod{
    \def\lay{\Ncnt} % alias of index of current layer
    \pgfmathsetmacro\prev{int(\Ncnt-1)} % number of previous layer
    \foreach \i [evaluate={\c=int(\i==\N); \y=\N/2-\i-\c*\yshift;
                 \x=\lay; \n=\nstyle;
                 \index=(\i<\N?int(\i):"\Nstr[\n]");}] in {1,...,\N}{ % loop over nodes
      % NODES
      \node[node \n] (N\lay-\i) at (\x,\y) {};
      % CONNECTIONS
      \ifnumcomp{\lay}{>}{1}{ % connect to previous layer
        \foreach \j in {1,...,\Nnod[\prev]}{ % loop over nodes in previous layer
          \draw[white,line width=1,shorten >=1] (N\prev-\j) -- (N\lay-\i);
          \draw[connect] (N\prev-\j) -- (N\lay-\i);
        }
        \ifnum \lay=\Nnodlen
          \draw[connect] (N\lay-\i) --++ (0.5,0); % arrows in; % arrows out
          \draw[connect] (N\lay-\i) --++ (1.5,0) node[right, above , text = black] {$\strut\Cstr[\n]_{\index}$}; % arrows in; % arrows out
        \fi
      }{
        \draw[connect] (0.2,\y) -- (N\lay-\i) node[above, midway, text = black] {$\strut\Cstr[\n]_{\index}$}; % arrows in
      }
      
    }
    \path (N\lay-\N) --++ (0,1+\yshift) node[midway,scale=1] {$\vdots$}; % dots
  }
  % LABELS
 \node[rectangle,draw = myblue,text = black,fill = myblue!20,minimum width = 22, minimum height = 90] (r) at (2.81,-0.75) {Softmax};
\end{tikzpicture}
\caption{ Proposed low-complexity \ac{ML} design for beam prediction.}
\label{fig:NNModel}
\end{figure}
\fi

\section{Performance Evaluation}
This section details detaset generation, model training, complexity analysis, and performance evaluation over specified \acp{KPI}. For reproducibility of results, our simulation dataset and source code is publicly available~\cite{source}.
\ifOneCol
\begin{table}[!t]
\centering
\caption{List of simulation parameters.}
 \begin{tabular}{llr@{\,}l}\toprule
 \multicolumn{2}{l}{\textbf{Parameters}} & \multicolumn{2}{l}{\textbf{Values}} \\ \toprule
 No. of \ac{BS} antennas &$N_{\mathrm{T}}$ & $64$ \\ 
 \ac{UE} antenna configuration &$N_{\mathrm{R}}$& $8$ \\ 
 \ac{BS} codebook size &$F$ & $64$ \\
 \ac{BS} parent codebook size &$F_{\mathrm{p}}$ & $16$ \\ 
 Transmit power &$P$ & $30$ &dBm\\
 \ac{BS} antenna gain  & & $8$ &dBi \cite{3GPP5}\\ 
 \ac{UE} codebook size&$W$ & $8$ \\ 
 \ac{UE} noise figure &$N_{\mathrm{F}}$  & $6$ &dB \\ 
 \ac{UE} antenna gain  & & $5$ &dBi \cite{3GPP6}\\ 
  Center frequency &$f_{\mathrm{c}}$ & $28$ &GHz \\ 
  Bandwidth &$B$ & $100$ &MHz \\ 
  Sub-carrier spacing & & $120$ &kHz\\ 
  Cell radius & &$200$ &m\\
 \bottomrule
 \end{tabular}
\label{simParams}
\end{table}
\fi
\subsection{Dataset Generation and Model Training}\label{A}
For dataset collection, we utilize the \ac{EBS} approach in combination with \ac{HBS}. Our dataset consists of parent \ac{RSRP} measurements, i.e., \ac{RSRP}$^{\mathrm{p}}$ obtained via the traditional \ac{HBS} and are provided as input features to the \ac{ML} model. In addition, the offline training labels, i.e., optimal beam indices are obtained via the traditional \ac{EBS} \cite{3GPP7}. Table \ref{simParams} lists default simulation parameters. The location of the \ac{UE} is drawn based on a uniform spatial distribution in the cell coverage area. The noise power $\sigma^{2}$ is computed as $(-174+10\text{log}_{10}B+N_{\text{F}})$~dBm and the path loss is given as $(20\text{log}_{10}d+20\text{log}_{10}f_{\mathrm{c}}-147.56)$~dB, where $d$ indicates distance. Finally, the channel model is considered as a \ac{CDL} model~\cite{3GPP5}. Further, to investigate the generalization capabilities of our \ac{ML} model, we consider following scenarios with different combinations of channel profiles \cite{3GPP7}.
    \begin{itemize}
    \item {Scenario ${1}$}: The \ac{ML} model is trained based on a training dataset constructed by utilizing the \ac{CDL}-D channel profile and performs inference on the \ac{UE} with same channel profile but with unknown location.
    \item {Scenario ${2}$}: The \ac{ML} model is trained based on a training dataset constructed by utilizing the \ac{CDL}-D channel profile and performs inference on a \ac{UE} with the \ac{CDL}-E channel profile and with unknown location.
    \item {Scenario ${3}$}: The \ac{ML} model is trained on the mixed dataset from above scenarios and performs inference on the \ac{UE} of both channel profiles but with unknown location.
\end{itemize}
Our dataset consists of $\num{25000}$ samples, where the training, validation, and testing data split is $70\%$, $10\%$, and $20\%$, respectively. Further, the \ac{ML} model is trained for $n_{\mathrm{e}}=100$ epochs, the model parameters are optimized by the Adam optimizer \cite{kingma2014adam} with the mean square error as loss function.

\ifOneCol
\else
\begin{table}[!t]
\centering
\caption{List of simulation parameters.}
 \begin{tabular}{llr@{\,}l}\toprule
 \multicolumn{2}{l}{\textbf{Parameters}} & \multicolumn{2}{l}{\textbf{Values}} \\ \toprule
 No. of \ac{BS} antennas &$N_{\mathrm{T}}$ & $64$ \\ 
 \ac{UE} antenna configuration &$N_{\mathrm{R}}$& $8$ \\ 
 \ac{BS} codebook size &$F$ & $64$ \\
 \ac{BS} parent codebook size &$F_{\mathrm{p}}$ & $16$ \\ 
 Transmit power &$P$ & $30$ &dBm\\
 \ac{BS} antenna gain  & & $8$ &dBi \cite{3GPP5}\\ 
 \ac{UE} codebook size&$W$ & $8$ \\ 
 \ac{UE} noise figure &$N_{\mathrm{F}}$  & $6$ &dB \\ 
 \ac{UE} antenna gain  & & $5$ &dBi \cite{3GPP6}\\ 
  Center frequency &$f_{\mathrm{c}}$ & $28$ &GHz \\ 
  Bandwidth &$B$ & $100$ &MHz \\ 
  Sub-carrier spacing & & $120$ &kHz\\ 
  Cell radius & &$200$ &m\\
 \bottomrule
 \end{tabular}
\label{simParams}
\end{table}
\fi

\subsection{Key Performance Indicators}
For performance evaluation in terms of beam measurement overhead, the \ac{KPI} is selected as reference signalling overhead reduction (\%) $1-\frac{N}{M}$, where $N$ is the number of beams (\acp{SSB}) required as input by the \ac{ML} model and $M$ is the total number of beams to be predicted \cite{3GPP7}. For beam prediction accuracy, the \ac{KPI} Top-$K$ (\%) is defined as the percentage that the truly optimal genie-aided transmit beam is among the $K$ best beams predicted by the ML model and the beam prediction error (\%) is calculated as $1-\text{beam prediction accuracy}$. Here, the Top-$1$ genie-aided transmit beam is obtained via \ac{EBS} \cite{3GPP7}. Further, the beam prediction accuracy is also evaluated in terms of achieved average \ac{RSRP}. Finally, for complexity analysis, we compare the model complexity in terms of number of trainable parameters and number of \acp{FLOP}.

\subsection{Complexity Analysis}
An important measure of \ac{ML} model complexity is the number of trainable parameters ($n_{\mathrm{l}}$), which for an \ac{FC}-\ac{NN} layer with $n_{\mathrm{i}}$ inputs and $n_{\mathrm{o}}$ outputs can be computed as $n_{\mathrm{l}} = (n_{\mathrm{i}}+1)n_{\mathrm{o}}$. Consequently, for proposed model the number of parameters are $n_{\mathrm{l}} = (F_{p}+1)F = 1088$. Further, the number of trainable  parameters for a convolutional layer can be obtained as $n_{\mathrm{l}} = n_{\mathrm{f}}(f_{\mathrm{h}}f_{\mathrm{w}}f_{\mathrm{d}}+1)$, where $n_{\mathrm{f}}, f_{\mathrm{h}}, f_{\mathrm{w}},$ and $f_{\mathrm{d}}$ indicate the number of filters, filter height, width, and depth, respectively. We evaluate the complexity in terms of model size with $32$-bit precision. Table \ref{tab2} indicates that due to a smaller number of trainable parameters the proposed model has the smallest size as compared to other models. 
\ifOneCol
\begin{table}[t]
\centering
\caption{Computational complexity comparison.}
\begin{tabular}{p{3cm}C{3cm}C{3cm}C{3cm}} \toprule
& \textbf{No. of Trainable Parameters}& \textbf{Model Size} (Mbits) & \textbf{No. of \acp{FLOP}} \\ \toprule
  \ac{FC}-\ac{NN} in \cite{9013296}  & \num{17728}                      & \num{0.5}             & \num[round-precision=3,round-mode=figures,scientific-notation=true]{17728} \\
  \ac{CNN} in \cite{9314253}         & \num{352034}\phantom{0}          & \num{11.2}\phantom{0} & \num[round-precision=3,round-mode=figures,scientific-notation=true]{1370000} \\
  \ac{CNN} in \cite{9492142}         & \num{67008}                      & \num{2.1}             &  \num[round-precision=3,round-mode=figures,scientific-notation=true]{332000} \\
  \ac{CNN} in \cite{3GPP4}           & \num{739073}\phantom{0}          & \num{23.6}\phantom{0} & \num[round-precision=3,round-mode=figures,scientific-notation=true]{47300000}\\
  Proposed model                     & \phantom{0}\num{1088}            & \phantom{0}\num{0.04} &\num[round-precision=3,round-mode=figures,scientific-notation=true]{1088}\\
  \bottomrule
\end{tabular}
\label{tab2}
\end{table}
\else
\begin{table}[t]
\centering
\caption{Computational complexity comparison.}
\begin{tabular}{p{1.9cm}C{2cm}C{1.4cm}C{1.4cm}} \toprule
& \textbf{No. of Trainable Parameters}& \textbf{Model Size} (Mbits) & \textbf{No. of \acp{FLOP}} \\ \toprule
  \ac{FC}-\ac{NN} in \cite{9013296}  & \num{17728}                      & \num{0.5}             & \num[round-precision=3,round-mode=figures,scientific-notation=true]{17728} \\
  \ac{CNN} in \cite{9314253}         & \num{352034}\phantom{0}          & \num{11.2}\phantom{0} & \num[round-precision=3,round-mode=figures,scientific-notation=true]{1370000} \\
  \ac{CNN} in \cite{9492142}         & \num{67008}                      & \num{2.1}             &  \num[round-precision=3,round-mode=figures,scientific-notation=true]{332000} \\
  \ac{CNN} in \cite{3GPP4}           & \num{739073}\phantom{0}          & \num{23.6}\phantom{0} & \num[round-precision=3,round-mode=figures,scientific-notation=true]{47300000}\\
  Proposed model                     & \phantom{0}\num{1088}            & \phantom{0}\num{0.04} &\num[round-precision=3,round-mode=figures,scientific-notation=true]{1088}\\
  \bottomrule
\end{tabular}
\label{tab2}
\end{table}
\fi

The time complexity of our proposed \ac{ML} model is compared in terms of number of required \acp{FLOP} using Big-$\mathcal{O}$ notation. During training, the \ac{ML} model performs forward and backward pass and it is useful to analyze the training and inference time complexity. In both forward and backward pass, the trainable parameters of a layer with $w$ nodes are updated by a matrix-vector multiplication resulting in a time complexity of $\mathcal{O}(w^{2})$ \acp{FLOP}. Furthermore, considering an \ac{NN} with $l$ layers, $w$ nodes per layer, and training the network with $n_{\mathrm{d}}$ data samples, and for $n_{\mathrm{e}}$ epochs requires $\mathcal{O}(n_{\mathrm{e}}n_{\mathrm{d}}lw^{2})$ \acp{FLOP} during training, while the inference requires only $\mathcal{O}(n_{\mathrm{d}}lw^{2})$ \acp{FLOP} as only forward pass is performed during inference. Similarly, the time complexity of a \ac{CNN}, with $l_{\mathrm{c}}$ convolutional and $l$ \ac{FC} layers during training is $\mathcal{O}(n_{\mathrm{e}}n_{\mathrm{d}}(n_{\mathrm{f}}l_{\mathrm{c}}i_{\mathrm{h}}i_{\mathrm{w}}(f_{\mathrm{h}}f_{\mathrm{w}}f_{\mathrm{d}}))+lw^{2})$. Here, in addition to the parameters defined above, $i_{\mathrm{h}}$~and~$i_{\mathrm{w}}$ indicate input height and width, respectively. Further, the inference time complexity is then given as $\mathcal{O}(n_{\mathrm{d}}(n_{\mathrm{f}}l_{\mathrm{c}}i_{\mathrm{h}}i_{\mathrm{w}}(f_{\mathrm{h}}f_{\mathrm{w}}f_{\mathrm{d}}))+lw^{2})$. 

Table \ref{tab2} summarizes the complexity comparison with the state of the art. For a fair comparison, the number of estimated \acp{FLOP} are for one epoch and one data sample, i.e., $n_{\mathrm{e}}=n_{\mathrm{d}}=1$. Here, it can be seen that the proposed \ac{ML} model achieves significantly lower computational complexity and benefits from lower power consumption. Further, the execution of the proposed \ac{ML} model on an Intel i7-1185G7 processor indicates that the training time per epoch and per data sample is $\SI{9}{\micro\s}$, which allows efficient and less time consuming model retraining. Besides, the execution time for each prediction is around $\SI{2}{\micro\s}$ allowing faster beam prediction.

\subsection{Simulation Results}
For performance evaluation, in addition to the  two-level \ac{HBS}, \ac{CNN} from \cite{9314253}, and the \ac{FC}-\ac{NN} from \cite{9013296}, the \ac{EBS} based beam selection is selected as a baseline for comparison \cite{3GPP7}. During inference the input to all \ac{ML} models are the \ac{RSRP}$^{\mathrm{p}}$ measurements of the parent beams and the outputs are the predicted probabilities of each child beam being the best. 

In terms of beam measurement overhead, the baseline \ac{EBS} requires $64$ beam measurements, resulting in $100\%$ beam measurement overhead. \Ac{HBS} requires $16$ parent and $4$ child beam measurements, resulting in beam measurement overhead of $32\%$. For the ML models, during inference, the measurement overhead depends on the value of $K \in \{4,2,1\}$, reflecting the necessity of probing the remaining $K$ beams for final selection, resulting in beam measurement overhead of around $32\%$, $28\%$, and $25\%$, respectively, as shown in Fig. \ref{fig:accVSoh}. In terms of beam prediction error for $K = 1$, our proposed approach reduces the error by around $2$, $1.4$, and $1$ percentage points as compared to \ac{HBS}, \cite{9013296}, and \cite{9314253}, respectively. %For $K = 2$ and $K = 4$ our proposed approach still performs equally good as \cite{9013296} and \cite{9314253}, while outperforming the traditional multilevel \ac{HBS}. 
Similar observations can be made from Fig. \ref{fig:rsrp}, where the performance is compared in terms of the average \ac{RSRP}. Here it can be noticed that the mean \ac{RSRP} achieved by all \ac{ML} approaches is well within a $0.15$ dB margin of the genie-aided (\ac{EBS}) transmit beam. However, it is worth mentioning that the \ac{HBS} achieves similar performance at the cost of increased latency.
\ifOneCol
\begin{figure}[t]
	\centering
	\includegraphics{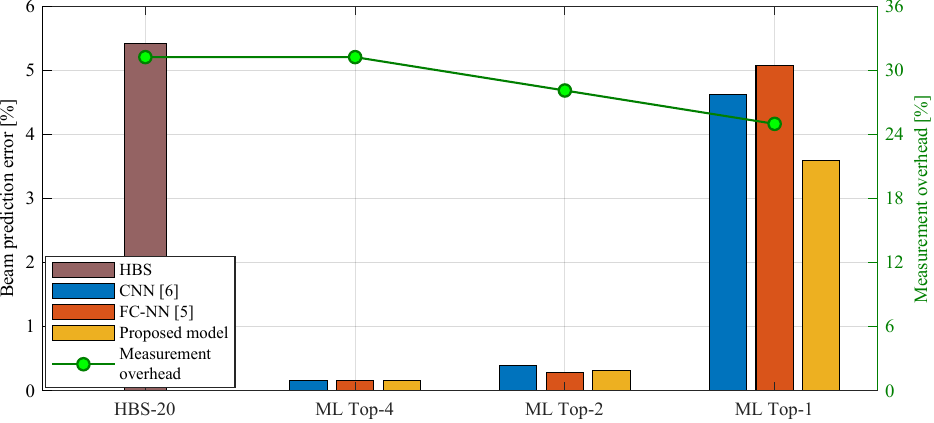}
	\caption{Beam prediction error and beam measurement overhead for scenario~1.}
	\label{fig:accVSoh}
\end{figure}
\else
\begin{figure}[t]
	\centering
	\includegraphics{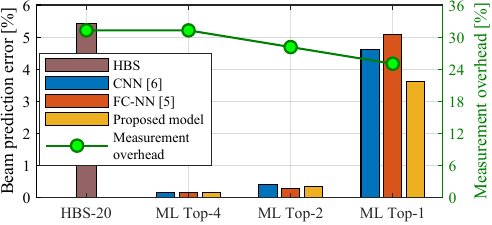}
	\caption{Beam prediction error and beam measurement overhead for scenario~1.}
	\label{fig:accVSoh}
\end{figure}
\fi
\ifOneCol
\begin{figure}[t]
	\centering
	\includegraphics{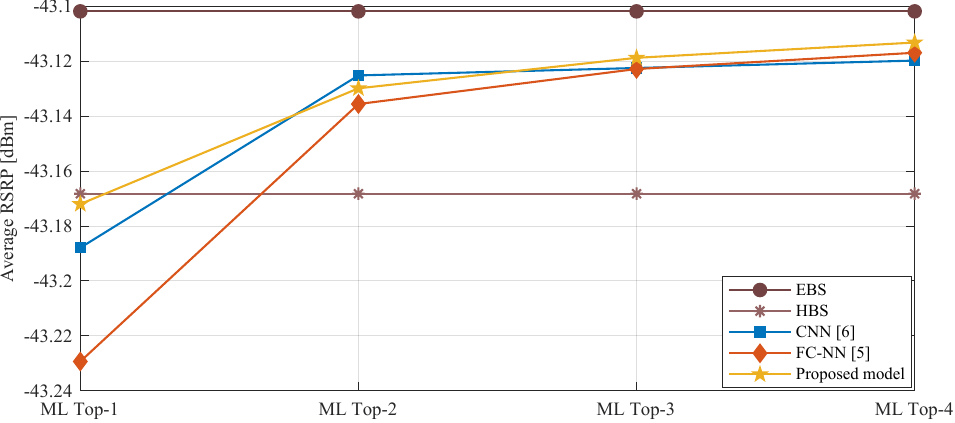}
	\caption{Comparison in terms of achieved average \ac{RSRP} [dBm] for scenario~1.}
	\label{fig:rsrp}
\end{figure}
\else
\begin{figure}[b]
	\centering
	\includegraphics{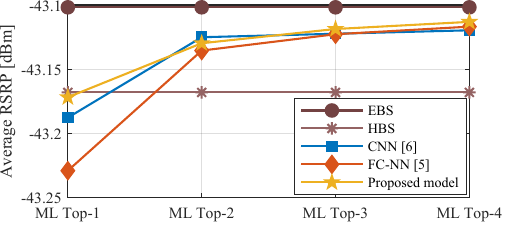}
	\caption{Comparison in terms of achieved average \ac{RSRP} [dBm] for scenario~1.}
	\label{fig:rsrp}
\end{figure}
\fi
\ifOneCol
\begin{figure}[b]
	\centering
	\includegraphics{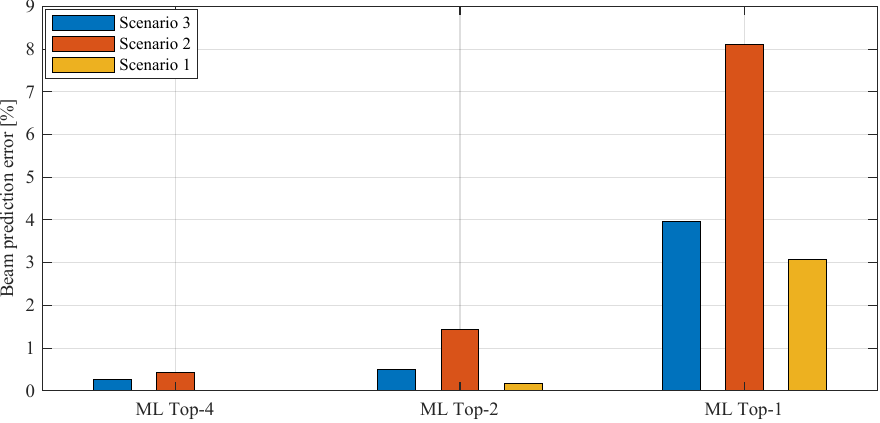}
	\caption{Generalization capabilities of proposed beam prediction model over \ac{3GPP} specified scenarios discussed in Section \ref{A}.}
	\label{fig:scenario}
\end{figure}
\fi

Fig. \ref{fig:scenario} showcases the generalization capabilities of our proposed model over three different scenarios as discussed in Section \ref{A}. We observe that for \ac{ML} Top-1 the prediction error of the model increases by around $5$ percentage points for scenario $2$, due to different channel profiles used in training and testing. Further, the error can be reduced when the model is trained on a mixed data set from different channel profiles, i.e., scenario~$3$. However, the error in scenario $3$ is still around $1.2$ percentage points higher as compared to scenario $1$. An important observation made here is that training a model for a large number of scenarios results in reduced inference performance for a specific scenario. Thus, there exists a trade-off between \ac{ML} model accuracy performance and its generalization capabilities.     
\ifOneCol
\else
\begin{figure}[t]
	\centering
	\includegraphics{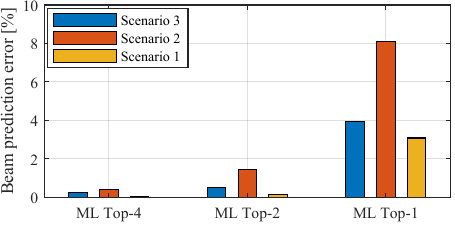}
	\caption{Generalization capabilities of proposed beam prediction model over \ac{3GPP} specified scenarios discussed in Section \ref{A}.}
	\label{fig:scenario}
\end{figure}
\fi

\section{Conclusion}
This letter proposes an \ac{ML}-based beam prediction design that reduces the reference signaling overhead and predicts the transmit beam with higher accuracy and much lower computational complexity as compared to the state-of-the-art. Specifically, we formulated the beam prediction problem as a multiclass-classification task and proposed a low-complexity \ac{ML} design to learn the spatial angular correlation between parent and child beams to predict the optimal beam. Due to lower computational complexity, the proposed model reduces the power consumption at the \ac{UE} and the beam prediction time making it suitable for faster beam prediction. Further, through simulation results, we showed that there exists a trade-off between \ac{ML} model performance and its generalization capabilities. These \ac{3GPP} compliant evaluation results indicate the feasibility of \ac{ML}-based \ac{mmWave} beam prediction for \ac{5G}-Advanced \ac{NR} and beyond \ac{5G} communication networks.

\bibliographystyle{IEEEtran}
\bibliography{IEEEabrv,./Bibliography.bib}
%TC:endignore

\end{document}